\documentclass[fleqn]{article}
\usepackage{amsmath,amssymb,epsfig}
\usepackage{bm}
\usepackage{amsfonts}
\usepackage{slashed}
\usepackage{graphicx,color}
\usepackage{xspace}
\usepackage{rotating,amsmath,amssymb}
\usepackage{epsfig}\parskip 5pt plus 1pt
\newcommand{\be}{\begin{equation}}
\newcommand{\ee}{\end{equation}}
\newcommand{\bea}{\begin{eqnarray}}
\newcommand{\eea}{\end{eqnarray}}
\newcommand{\ba}{\begin{array}}
\newcommand{\ea}{\end{array}}

\def\simge{\mathrel{%
   \rlap{\raise 0.511ex \hbox{$>$}}{\lower 0.511ex \hbox{$\sim$}}}}
\def\simle{\mathrel{
   \rlap{\raise 0.511ex \hbox{$<$}}{\lower 0.511ex \hbox{$\sim$}}}}

\def\nue{\ensuremath{\nu_{e}}\xspace}
\def\nubare{\ensuremath{\overline{\nu}_{e}}}

\def\numu{\ensuremath{\nu_{\mu}}\xspace}

\def\nutau{\ensuremath{\nu_{\tau}\xspace}}

\newcommand{\dmot}{\ensuremath{\Delta m^2_{12}\xspace}}
\newcommand{\dmtt}{\ensuremath{\Delta m^2_{23} \xspace}}

\newcommand{\thetaot}{\ensuremath{\theta_{13}}\xspace}
\newcommand{\thetaotw}{\ensuremath{\theta_{12}}\xspace}
\newcommand{\thetatt}{\ensuremath{\theta_{23}}\xspace}
\newcommand{\numunue}{\ensuremath{\nu_\mu \rightarrow \nu_e}\xspace}

\def\ri{\right}
\def\lf{\left}
\def\si{\sigma}
\def\te{\theta}

\def\lf{\left}

\def\non{\nonumber}

\def\ri{\right}
\def\al{\alpha}

\def\te{\theta}
\def\si{\sigma}

\def\1{{_{1}}}\def\2{{_{2}}}

\newcommand{\ide}{1\hspace{-1mm}{\rm I}}

\def\noHe0{:\;\!\!\;\!\!:H_e(0):\;\!\!\;\!\!:}
\def\noHm0{:\;\!\!\;\!\!:H_\mu(0):\;\!\!\;\!\!:}

\def\vect#1{{\bm #1}}

\def\lf{\left}

\def\non{\nonumber}

\def\ri{\right}

\def\al{\alpha}

\def\te{\theta}

\def\si{\sigma}

\def\1{{_{1}}}\def\2{{_{2}}}
\begin{document}
\thispagestyle{empty}
\vspace*{1cm}
\begin{center}
{\Large{\bf An infrared origin of leptonic mixing and its test at DeepCore} }\\
\vspace{.5cm} 

F.~Terranova$^{\rm a,b}$ \\
\vspace*{1cm}
$^{\rm a}$ I.N.F.N., Laboratori Nazionali di Frascati,
Frascati (Rome), Italy \\
$^{\rm b}$ Dipartimento di Fisica, Universit\`a di Milano Bicocca, 
Milano, Italy \\ 
\end{center}

\vspace{.3cm}
\begin{abstract}
\noindent
Fermion mixing is generally believed to be a low-energy manifestation
of an underlying theory whose energy scale is much larger than the
electroweak scale. In this paper we investigate the possibility that
the parameters describing lepton mixing actually arise from the
low-energy behavior of the neutrino interacting fields.  In
particular, we conjecture that the measured value of the mixing angles
for a given process depends on the number of unobservable flavor states
at the energy of the process. We provide a covariant implementation of
such conjecture, draw its consequences in a two neutrino family
approximation and compare these findings with current experimental
data. Finally we show that this infrared origin of mixing will be
manifest at the Ice Cube DeepCore array, which measures atmospheric
oscillations at energies much larger than the tau lepton mass; it will
hence be experimentally tested in a short time scale.
\end{abstract}

\vspace*{\stretch{2}}
\begin{flushleft}
\end{flushleft}

\newpage

\section{Introduction}
\label{introduction}

Mixing of elementary fermions is very well established from the
experimental point of view~\cite{PDG} although the origin of the
parameters that describe the quark~\cite{CKM} and lepton~\cite{PMNS}
mixing matrix remains a deep mystery in modern particle physics. Since
the Standard Model (SM) is probably an effective theory up to some
energy scale, above which new physics has to be accounted for, it is
likely that flavor mixing has an ultraviolet origin, too. In
fact, it is commonly believed that flavor mixing is a manifestation of
an underlying theory whose energy scale resides well above the
electroweak scale $v=(G_F \sqrt{2})^{-1/2}$. In this
framework, the peculiar structure of the mixing
matrices and, in particular, the striking difference between quark and
lepton mixing can shed light on the symmetries of the underling
theory~\cite{Altarelli:2010gt}, even if its energy scale is
unattainable by high-energy
accelerators~\cite{Fritzsch:1999ee,Albright:2006cw}.

Though the ultraviolet origin remains the most plausible explanation
of fermion mixing, in this paper we follow a different path and we
consider that mixing might arise as a consequence of the low energy
behavior of the interacting fermion fields. More precisely, we
decouple the problem of the origin of the mass from the problem of
flavor mixing assuming that the former is due to an ultraviolet
mechanism (as the Higgs mechanism with diagonal Yukawa couplings)
while the latter arises at energies $\ll v$ (``infrared origin'').

We are driven in such consideration by a few reasons. Firstly, an
infrared origin can naturally produce opposite mixing structures
between quark and leptons, since the low-energy behavior of the
corresponding interacting fields is very different. It can
also explain the persistent difficulties in linking the concept of
``flavor neutrino states'' to the standard properties of Fock states
in quantum field theory (QFT)~\cite{book_blasone}. Even more,
experimental neutrino data show intriguing features when interpreted
not only as a function of the energy and source-to-detector baseline
but also of the number of kinematic thresholds ($N$) for lepton production
that the neutrino is able to cross. 
   
The core of this paper is a conjecture that links the number of
unobservable flavor states for a given process (3-$N$ if we assume the
standard 3-family scenario) to the measured value of the mixing angle
for such process. This conjecture is discussed and stated in the
standard QFT framework in Sec.~\ref{sec:conjecture}. A consistent
implementation of the conjecture, which allows extracting specific
predictions on neutrino mixing, is developed in
Sec.~\ref{sec:flavor_states}. Phenomenological implications,
especially for the Ice Cube DeepCore array~\cite{deepcore_det}, and
comparison with existing data are presented in Sec.~\ref{sec:data} and
summarized in the Conclusions (Sec.~\ref{sec:conclusions}).
   
\section{Flavor projectors}
\label{sec:conjecture}

The Standard Model and the Minimally Extended SM\footnote{I.e. the SM
  supplemented with right-handed neutrinos that are singlets under the
  electroweak gauge group and allow for a Dirac neutrino mass through
  the Higgs mechanism~\cite{Giunti:2007ry}.}  do not make predictions
  on the values of the elementary fermion masses and their mixing
  parameters (angles and CP violating phases). However, they entangle
  the problem of fermion mass generation with mixing through the Higgs
  mechanism. Indeed, the  Higgs-fermion Yukawa
  Lagrangian of the Minimally Extended SM reads, in unitary gauge:

\bea 
\mathcal{L}_{H,F} &=& - \frac{v+H}{\sqrt{2}} \left[ 
\sum_{\alpha,\beta=d,s,b} Y'^D_{\alpha \beta} \overline{ q'^D_{\alpha
L} } q'^D_{\beta R} + \sum_{\alpha,\beta=u,c,t} Y'^U_{\alpha \beta}
\overline{ q'^U_{\alpha L} } q'^U_{\beta R} +  \right. \nonumber \\
&& \left. \sum_{\alpha,\beta=e,\mu,\tau}
Y'^D_{\alpha \beta} \overline{ q'^D_{\alpha L} } q'^D_{\beta R} +
\sum_{\alpha,\beta=\nue,\numu,\nutau}
Y'^D_{\alpha \beta} \overline{ q'^U_{\alpha L} } q'^U_{\beta R} \right]
+ H.c.
\eea

\noindent
where $U$ and $D$ labels the up-type and down-type fermions $q'$; L, R
their chirality and $H$ the Higgs field; the Yukawa matrices
$Y'_{\alpha \beta}$ are generic $3\times3$ matrices that can be
diagonalized by biunitary transformations.  No experiment has direct
access to the Higgs-fermion couplings and, actually, the Higgs sector
has not been established, yet; hence the only source of observables to
discriminate weak flavor eigenstates from mass eigenstates remains
charged-current (CC) interactions. In the SM (for quarks) and in the
Minimally Extended SM (for quarks and leptons), the fact that CC
interactions are the only source of flavor projectors is both due to
flavor independence of e.m. + strong interactions and to the
effectiveness of the GIM mechanism.
In CC, the matrices that diagonalize the left-handed and right-handed,
up-type and down-type fermions ($V_L^{U\dagger}$, $V_R^{U\dagger}$,
$V_L^{D\dagger}$ and $V_L^{D\dagger}$) appear only in the combination
$U= V_L^{U\dagger} V_L^{D}$, where $U$ indicates the CKM matrix for
quark and the Pontecorvo-Maki-Nakagawa-Sakata (PMNS) matrix for
leptons. For the neutral current (NC) part of the Lagrangian, the
corresponding combinations are $ V_L^{U\dagger} V_L^{U} =
V_D^{U\dagger} V_D^{U} = V_R^{U\dagger} V_R^{U} = V_R^{U\dagger}
V_R^{U} = \ide $. As a consequence, in the NC sector, the CKM and PMNS
matrices are unphysical and the theory is symmetric under flavor
exchange. It means, for instance, that a NC scattering amplitude is
the same for any transformation $\alpha \rightarrow \beta$ that
changes the flavor label of the neutrino ($\alpha=e, \mu, \tau$).
This is equivalent to the usual statement that NC processes are
flavor-independent.

Alternatively, we can re-state the previous result saying that, in the
Minimally Extended SM, the removal of all flavor projectors (CC
interactions) restores flavor symmetry since the corresponding quantum
number becomes unphysical. This is the statement we want to broaden
beyond the limit of applicability of the Minimally Extended SM.

\begin{figure}
  \begin{center}
    \includegraphics{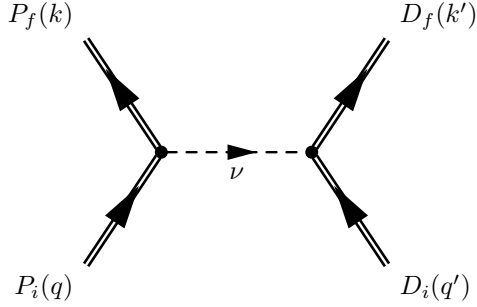}
  \end{center}
\caption{Feynman diagram that describes production, propagation 
           and detection of a neutrino as a single process.}
  \label{fig:diagram}
\end{figure}

In order to do so, we assume without loss of
generality~\cite{Akhmedov:2010ms,Beuthe:2001rc} that the process under
consideration involves one initial state and one final state particle
besides the neutrino and we employ the standard QFT formalism: hence,
production, propagation and detection are considered simultaneously
through the diagram of Fig.\ref{fig:diagram}.
Following~\cite{Akhmedov:2010ms}, we therefore define
the states describing the particles accompanying neutrino production and 
detection as:
\be
|P_i\rangle =\int\! [d q] 
\,f_{Pi}(\vec{q},\vec{Q})\,|P_i,\vec{q}\rangle\,,\qquad
|P_f\rangle =\int\! [d k]
\,f_{Pf}(\vec{k},\vec{K})\,|P_f,\vec{k}\rangle\,,
\label{eq:state2}
\ee
and
\be
|D_i\rangle =\int\! [d q']
\,f_{Di}(\vec{q}',\vec{Q}')\,|D_i,\vec{q}'\rangle\,,\qquad
|D_f\rangle =\int\! [dk']
\,f_{Df}(\vec{k}',\vec{K}')\,|D_f,\vec{k}'\rangle\,.
\label{eq:state3}
\ee 
For any momentum label of Fig.~\ref{fig:diagram}, we shortened the
notation defining 
\be 
[dp]\equiv\frac{d^3
  p}{(2\pi)^3\sqrt{2E_A(\vec{p})}}\,.
\label{eq:not1}
\ee
In the formulas above, $|A,\vec{p}\rangle$ is the one-particle momentum eigenstate
corresponding to momentum $\vec{p}$ and energy $E_A(\vec{p})$, and
$f_A(\vec{p},\vec{P})$ is the momentum distribution function with 
$\vec{P}$ as mean momentum. Through the use of $f_A(\vec{p},\vec{P})$
we are aiming at computing the amplitude of
Fig.~\ref{fig:diagram} employing an external wave packet
approach~\cite{Beuthe:2001rc,sachs,Giunti:1993se}. The main advantage
of this approach is that it allows for a perturbative evaluation of
the transition amplitude without resorting to the definition of
``flavor states''. To see this, we first note that the amplitude of
the Feynman diagram of Fig.~\ref{fig:diagram} is given by:
\be
i{\cal A}_{\alpha\beta}=\langle P_f \,D_f|\hat{T}\exp\Big[-i\int \!d^4x 
\,{\cal H}_I(x)\Big]-\ide |P_i \,D_i\rangle\,,
\label{eq:amp2}
\ee 
where $\hat{T}$ is the time ordering operator and ${\cal H}_I(x)$ is
the CC weak interaction Hamiltonian, i.e. the part of the Hamiltonian
that generates the flavor projectors for the process under
consideration.

Eq.\ref{eq:amp2} can be written explicitly as~\cite{Akhmedov:2010ms}:
\bea
i{\cal A}_{\alpha\beta}&=&\sum_j U_{\alpha j}^* U_{\beta j}
\int\! [dq]
\,f_{Pi}(\vec{q},\vec{Q}) \int\! [dk]
\,f^*_{Pf}(\vec{k},\vec{K}) 
\nonumber \\
& &
\times \int\! [dq']
\, f_{Di}(\vec{q}',\vec{Q}') \int\! [dk']
\,f^*_{Df}(\vec{k}',\vec{K}')
\,i{\cal A}^{p.w.}_j(q,k;q',k')\,.
\label{eq:amp3}  
\eea

The formalism is built in such a way that all intermediate states over
which the sum is running are actually neutrino mass eigenstates, not
flavor eigenstates. On the technical side, the quantity ${\cal
  A}^{p.w.}_j(q,k;q',k')$ is the plane-wave amplitude of the process
with the $j$th neutrino mass eigenstate propagating between the source
and the detector:
\begin{align}
i{\cal A}^{p.w.}_j(q,k;q',k')=&
\int d^4 x_1 \!\int d^4 x_2 \,
\tilde{M}_D(q',k')\, e^{-i(q'-k')(x_2-x_D)}
\nonumber \\
& \times
i\!\int\!\frac{d^4 p}{(2\pi)^4} \frac{\slashed{p}+m_j}
{p^2-m_j^2+i\epsilon}\, e^{-i p (x_2-x_1)}
\cdot \tilde{M}_P(q,k)\, e^{-i(q-k)(x_1-x_P)} \,. 
\label{eq:amp4}
\end{align}

\noindent
Here $x_1$ and $x_2$ are the 4-coordinates of the neutrino production
and detection points. Integration over these coordinates brings the
delta functions that impose energy and momentum conservation at the
source and at the detector. The quantities $\tilde{M}_P(q,k)$ and
$\tilde{M}_D(q',k')$ are the plane-wave amplitudes of the processes
$P_i\to P_f+\nu_j$ and $D_i+\nu_j\to D_f$, respectively, with the
neutrino spinors $\bar{u}_j(p,s)$ and $u_j(p,s)$ excluded ($s$ is the
neutrino spin variable).

As noted in \cite{Akhmedov:2010ms,Giunti:2006fr}, the external wave
packet approach allows for a consistent derivation of the Pontecorvo
oscillation formula and sidesteps the definition of ``flavor states''
that poses some formal hurdles~\cite{Giunti:2003dg,Blasone:2006jx}.
On the other hand, Eq.~\ref{eq:amp3} obscures a subtle feature that
can be of interest to understand the origin of lepton mixing. For sake
of definiteness, let us consider the manifold of the amplitudes of
Eq.~\ref{eq:amp3} for a pure CC process, where the final state is a
lepton of mass $m_\alpha$ ($\alpha=e,\mu,\tau$) accompanied by $n$
additional particles of mass $M_i$. Since the momentum spread of the
initial state particles is generally smaller than the mass of the
heavier lepton, we can classify the manifold using the mean momenta of
the $f_A(\vec{p},\vec{P})$ functions. In the laboratory
frame\footnote{This assumption is done without loss of generality,
  since the number of kinematic thresholds that are crossed in a given
  process is a Lorentz invariant quantity.}, i.e. in the reference
frame where the neutrino target A is at rest ($\vec{Q}'=0,
E_{\vec{Q}'}= M_A$), the manifold can be expressed as
\begin{equation}
{\cal A}_{\alpha\beta} \simeq {\cal A}_{\alpha\beta}(\vec{Q},\vec{K},\vec{K}')
\label{eq:manifold}
\end{equation}
and, for each value of $\alpha$ and $\beta$, it can be classified in
nearly~\cite{why_nearly} disconnected sets checking whether, for a
given $\alpha$, the initial state momenta are above the kinematic threshold
for the production of the charged lepton $\beta$:
\begin{equation}
E_\nu^{thr} \simeq | \vec{Q}-\vec{K} | > 
\frac{( \sum_i M_i + m_\alpha )^2}{2 M_A} - \frac{M_A}{2}
\label{eq:threshold}
\end{equation}

For initial states where most of the kinematical thresholds are
forbidden, it is an experimental fact that mixing (toward unobservable
flavors) is naturally large\footnote{Mixing is observable if the
  oscillation phase is made large by an appropriate choice of the
  neutrino energy and baseline. Since we are assuming that the
  hierarchy of mass eigenstates, i.e. the values of $m_1$, $m_2$ and
  $m_3$, has an ultraviolet origin and it is decoupled from the values
  of the mixing angles, the oscillation frequency at the solar and
  atmospheric scales are the same as for the standard three-family
  interpretation ($\dmot \simeq 8 \times 10^{-5}$~eV$^2$ and $\dmtt
  \simeq 2.4 \times 10^{-3}$~eV$^2$)~\cite{Schwetz:2011qt}.}. This is the case of solar and
reactor neutrinos, where the initial state is $\nue$ ($\nubare$) and
the kinematic threshold for muon production is well beyond the
neutrino energy. Here, the effective mixing angle (\thetaotw in the
standard three family interpretation) is $\simeq 33^\circ$. Similarly,
it is the case of $\numu$ oscillations at the atmospheric scale, where
$\numu$ are mostly below the kinematic threshold for tau
production and the corresponding mixing angle is $\simeq \pi/4$. On
the contrary, \numunue oscillations at the atmospheric scale, where all kinematic thresholds
are available, turns out to be small. In the standard
framework (see Sec.\ref{sec:data} for a discussion), the latter is
interpreted as indication for a small mixing between the first and
third family ($\thetaot < 12^\circ$).

The QFT formulation of neutrino oscillations depicted above
(Eqs.~\ref{eq:state2}-\ref{eq:amp4}) is able to compute in a consistent
manner the manifold (\ref{eq:manifold}) because integration over $x_1$
and $x_2$ embeds the threshold constraint and ${\cal A}_{\alpha\beta}$
goes to zero every time the condition (\ref{eq:threshold}) is not
fulfilled.  In this framework, however, the connection between the
kinematic thresholds that are open to neutrinos and the size of the mixing
angles is purely accidental.
We hence put forward the following \\

\noindent {\bf Conjecture (A)}: Mixing is a process dependent
phenomenon, whose size depends on the number of flavor states $N$ that can
potentially be observed through the production of the corresponding
lepton. Running from below to above the kinematic thresholds, the
mixing parameters change and in the limit $E_\nu \ll E^{thr}$ they
  settle to restore flavor invariance
  for the appropriate Hamiltonian.

In order to state quantitatively this conjecture, we need to write
explicitly the Hamiltonian for interacting flavor fields. This task will
be carried out in Sec.~\ref{sec:flavor_states}.

\section{Flavor states}
\label{sec:flavor_states}

The description of neutrino oscillations based on the external wave
packet and just resorting on the concept of mass eigenstate is
motivated by the difficult interpretation of flavor neutrino states in
QFT. These states should be eigenstates of the flavor charge and
should be the quanta of the corresponding flavor fields, obeying the standard
anticommutation rules for Dirac fermions. Unfortunately the most obvious choice
for a definition of a flavor state and field turns out to be inconsistent.

In particular, one would choose the linear combination: 

\begin{equation}
| \nu_{\alpha} \rangle
=
\sum_{k} U_{\alpha k}^* \, | \nu_k \rangle
\qquad
(\alpha=e,\mu,\tau)
\,,
\label{902}
\end{equation}
to be the natural candidate for a flavor neutrino state,
where
$| \nu_k \rangle$
is the state of a neutrino with mass $m_k$,
which belongs to the Fock space of the quantized
massive neutrino field $\nu_k$.
Similarly, 
the left-handed flavor neutrino fields $\nu_{\alpha L}$,
with $\alpha=e,\mu,\tau$,
should be unitary linear combinations of the massive neutrino fields
$\nu_{kL}$,
\begin{equation}
\nu_{\alpha L}
=
\sum_{k=1}^3 U_{\alpha k} \, \nu_{kL}
\qquad
(\alpha=e,\mu,\tau)
\,,
\label{901}
\end{equation}
where $U$ is the leptonic mixing matrix.
As a matter of fact, (\ref{902}) is not a quantum of the flavor
field $\nu_{\alpha}$ \cite{Giunti:1992cb} except for the trivial case
of massless neutrinos.  This statement holds as far as we make the
quite natural assumption that flavor destruction (creation)
operators must be a linear combination of destruction (creation)
operators of massive states only. 

In fact, it has been
shown in~\cite{BV95} that a consistent definition can be achieved in a rather
straightforward manner, at least in a two family approximation.
Flavor fields can be properly defined if we derive them as {\it transformed}
fields from the mass fields. The transformations are:
\bea \label{PontecorvoMix}
\nu_e(x) &=& \nu_1(x)\cos\theta + \nu_2(x)\sin\theta\\ [1mm] \nonumber
\nu_{\mu}(x) &=& -\nu_1(x)\sin\theta + \nu_2(x)\cos\theta, \eea
The starting field are, therefore, the massive  free fields $\nu_1$ and $\nu_2$, whose Fourier expansions are:
\bea \nu_j(x) = \int \frac{d^3
\mathbf{k}}{(2\pi)^{3/2}}\sum_r\lf[u^r_{\mathbf{k},j}(x_0)\alpha^r_{\mathbf{k},j}
+ v^r_{-\mathbf{k},j}(x_0)\beta^{r \dagger}_{\mathbf{-k},j}\ri]
e^{i\mathbf{k}\cdot\mathbf{x}}, \qquad j=1,2,
\eea
with $u^r_{\mathbf{k},j}(x_0) = u^r_{\mathbf{k},j} e^{-i
\omega_{\mathbf{k},j} x_0}$,
$v^r_{-\mathbf{k},j}(x_0)=v^r_{-\mathbf{k},j}e^{i
\omega_{\mathbf{k},j} x_0}$, and
$\omega_{\mathbf{k},j}=\sqrt{\mathbf{k}^2+m_j^2}$. The operators
$\alpha^r_{\mathbf{k},j}$ and $\beta^{r}_{\mathbf{-k},j}$,
$j=1,2$, $r=1,2$ are the annihilation operators for the vacuum
state $|0\rangle_{1,2}=|0\rangle_1\otimes|0\rangle_2$:
$\alpha^r_{\mathbf{k},j}|0\rangle_{1,2},
\beta^{r}_{\mathbf{-k},j}|0\rangle_{1,2}=0$. The canonical
anticommutation relations are: $\{\nu_i^{\alpha}(x),
\nu_j^{\beta\dagger}(y)\}_{x_0=y_0}=\delta^3(\mathbf{x}-\mathbf{y})
\delta_{\alpha\beta}\delta_{ij}$
with $\alpha, \beta=1,\ldots,4$ and
$\{\alpha^r_{\mathbf{k},i},\alpha^{s\dagger}_{\mathbf{q},j}
\}=\delta_{\mathbf{k}\mathbf{q}}\delta_{rs}\delta_{ij}$; $\{
\beta^r_{\mathbf{k},i},\beta^{s\dagger}_{\mathbf{q},j}
\}=\delta_{\mathbf{k}\mathbf{q}}\delta_{rs}\delta_{ij}$, with
$i,j=1,2$. All other anticommutators are zero. The ortonormality
and completeness relations are: $u_{{\bf k},j}^{r\dagger }u_{{\bf
k},j}^{s} = v_{{\bf k},j}^{r\dagger }v_{{\bf k},j}^{s}= \delta
_{rs}$, $u_{{\bf k},j}^{r\dagger }v_{-{\bf k},j}^{s} = v_{-{\bf k}
,j}^{r\dagger }u_{{\bf k},j}^{s}=0$, $ \sum_{r}(u_{{\bf
k},j}^{r}u_{{\bf k},j}^{r\dagger }+v_{-{\bf k},j}^{r}v_{-{\bf
k},j}^{r\dagger }) = 1$.

Flavor fields can hence be constructed from the generator of the mixing
transformation (\ref{PontecorvoMix}):
\bea \label{MixingRel} \nu^{\alpha}_\si(x)&=&
G_{\theta}^{-1}(x_0)\nu^{\alpha}_j G_{\theta}(x_0) \,,
\qquad (\sigma, j)=(e,1),(\mu,2), \eea
with $G_{\theta}(x_0)$  given by:
\be \label{MixGen} G_{\theta}(x_0)=\exp\left[\theta\int
d^3\mathbf{x}\left(\nu_1^{\dagger}(x)\nu_2(x)-
\nu_2^{\dagger}(x)\nu_1(x)\right)\right], \ee
The flavor
annihilators can be defined as:
\begin{eqnarray}\label{flavannich}
\alpha _{{\bf k},\sigma}^{r}(x_0) &\equiv &G^{-1}_{\bf
\te}(x_0)\;\alpha
_{{\bf k},j}^{r}\;G_{\bf \te}(x_0),   \qquad (\sigma, j)=(e,1), (\mu,2)
\end{eqnarray}
and similar ones are defined for the antiparticle operators.
In turn, flavor fields can be rewritten in the form:
\bea \nu_{\sigma}(x) = \int \frac{d^3
\mathbf{k}}{(2\pi)^{3/2}}\sum_r\lf[u^r_{\mathbf{k},j}(x_0)
\alpha^r_{\mathbf{k},\sigma}(x_0) +
v^r_{-\mathbf{k},j}(x_0)\beta^{r \dagger}_{\mathbf{-k},\sigma}(x_0)\ri]e^{i\mathbf{k}\cdot\mathbf{x}},
 \qquad (\sigma, j)=(e,1),(\mu,2),
\eea
i.e. they can be expanded in the same bases as the fields $\nu_i$. 

In spite of the apparent simplicity, a rich and troublesome
non-perturbative structure emerges from this definition. In
particular, the vacuum of flavor states is orthogonal to the vacuum of
the free fields, i.e. two Hilbert spaces are unitarily
inequivalent~\cite{BV95,tesi_capolupo}. The formalism retrieves the
standard Pontecorvo formula for neutrino
oscillations~\cite{BHV98,Blasone:2002wp} and gives consistent results
in the evaluation of the production-detection vertices of
Fig.~\ref{fig:diagram}~\cite{Blasone:2006jx,BCTV2005} but the
demonstrations are highly non-trivial. Finally, the flavor vacuum is
not Lorentz invariant being explicitly time-dependent.  Thus, flavor
states cannot be interpreted in terms of irreducible representations
of the Poincar\'{e} group. 

It has been recently pointed out~\cite{Blasone:2010zn} that the
difficulty of dealing with time-dependent vacuum states can be
technically overcome considering the flavor fields as interacting
fields with an external non-abelian gauge field $A_\mu$. As a
consequence, the mixed fields can be treated formally as free fields,
avoiding in this way the problems with their interpretation in terms
of the Poincar\'{e} group. The authors of \cite{Blasone:2010zn} note
that the presence of $A_\mu$ enables us to define flavor neutrino
states which are simultaneous eigenstates of the flavor charges, of
the momentum operators and of a new Hamiltonian operator for the mixed
fields. They interpret the new Hamiltonian qualitatively as the energy
which can be extracted from flavor neutrinos through scattering
although the physical meaning of $A_\mu$ (called ``neutrino aether''
in~\cite{Blasone:2010zn}) is unclear. In the following, we reconsider
the results of \cite{Blasone:2010zn} showing that $A_\mu$ can be
interpreted as an effective field arising when non-observable flavor
states can potentially contribute to Fig.~\ref{fig:diagram} and that the
new Hamiltonian is appropriate to re-state quantitatively Conjecture
(A).

Following~\cite{Blasone:2010zn}, the gauge field $A_\mu$ can be built starting from
the Euler-Lagrange equations
\bea i \partial_0 \nu_e &=& (-i \vect{\al}\cdot\vect{\nabla} + \beta m_e)\nu_e
+ \beta m_{e\mu} \nu_{\mu} \\[2mm] i \partial_0 \nu_{\mu} &=& (-i
\vect{\al}\cdot\vect{\nabla} + \beta m_{\mu})\nu_{\mu} + \beta m_{e\mu}
\nu_e, \label{eq:euler-lagrange} \eea
that corresponds to the Lagrangian density for two mixed neutrino fields:
\bea\label{Lagrflav} {\cal L} &=& {\bar \nu}_e \lf( i
\not\!\partial -
  m_{e}\ri)\nu_e  +  {\bar \nu}_\mu \lf( i \not\!\partial -
  m_{\mu}\ri)\nu_\mu \,- \, m_{e \mu} \,\lf({\bar \nu}_{e}\,
\nu_{\mu} \,+\, {\bar \nu}_{\mu} \,\nu_{e}\ri). \eea
Here, $\alpha_i$, $i=1,2,3$ and $\beta$ are the Dirac matrices in a
given representation and the masses in the Lagrangian are $ m_{e} =
m_{1}\cos^{2}\theta + m_{2} \sin^{2}\theta$, $m_{\mu} =
m_{1}\sin^{2}\theta + m_{2} \cos^{2}\theta$, $m_{e\mu}
=(m_{2}-m_{1})\sin\theta \cos\theta\,$. As in Eq.~\ref{PontecorvoMix}, the angle $\theta$ is the
leptonic Cabibbo angle, i.e. the only parameter describing mixing in two-family
approximation.

We define the external gauge field as
 \bea \label{Conn}
A_{\mu} &\equiv &\frac{1}{2} A_{\mu}^a  \si_a \,=\,  n_{\mu} \delta m \,\frac{\sigma_1}{2}\in
SU(2), \qquad n^{\mu} \equiv (1,0,0,0)^T ,\eea
$\sigma_i$ being the Pauli matrices and $\delta m \equiv m_\mu - m_e $. 
The corresponding covariant derivative is 
\bea D_{\mu}= \partial_{\mu} + i\, g\, \beta\, A_{\mu}, \eea
where the coupling constant is now $g\equiv \tan 2\theta$. This derivative can be easily
connected to the Euler-Lagrange equation. If we choose as representation for the Dirac matrices
\bea \alpha_i=\lf(\ba{cc}0&\sigma_i\\\sigma_i&0\ea\ri),\qquad
\beta= \lf(\ba{cc}\ide&0\\0&-\ide\ea\ri),\eea
$\ide$ being the $2\times 2$ identity matrix, the Euler--Lagrange  equations  can be written as:
\bea iD_0 \nu_f= (-i\vect{\al}\cdot\vect{\nabla} + \beta M_d)\nu_f,\eea
where $\nu_f= (\nu_e, \nu_\mu )^T$ is the flavor doublet,
$M_d={\rm diag}(m_e,m_{\mu})$ is a diagonal mass matrix and the covariant derivative is defined as
\bea \label{covdevmix}
D_0 \equiv \partial _0 + i\, m_{e\mu}\,\beta\,\sigma_1,  \eea
where $m_{e\mu}=\frac{1}{2}\,\tan 2\theta \,\delta m$. It thus follows
that the Lagrangian density (\ref{Lagrflav}) has the form of a doublet
of Dirac fields in interaction with an external Yang-Mills field:
\bea \label{LagrflavCov}{\cal L} = {\bar \nu}_f
(i\gamma^{\mu}D_{\mu} - M_d) \nu_f. 
\eea
As expected, the strength of the Yang-Mills field $g=\tan 2\theta$
vanishes for $\theta \rightarrow 0$ while the theory becomes
non-perturbative for $\theta \rightarrow \pi/4$.

It can be shown~\cite{Blasone:2010zn} that quantization of this theory
brings to flavor states that are eigenstates of flavor charges, of the three-momentum operators and of
a new Hamiltonian that follows from the energy-momentum tensor $\widetilde{T}^{\mu \nu}$ of the Lagrangian 
(\ref{LagrflavCov}). It is:
\bea \non \widetilde{H}(x_0) &=& \int d^3 \mathbf{x}\,
\widetilde{T}^{00}=  \int d^3 \mathbf{x} \,{\bar \nu}_f\lf( i  \gamma_0 D_0
- i \gamma^\mu D_\mu + M_d \ri)\nu_f
\\ \non
&=& \int d^3 \mathbf{x}\, \nu_e^{\dagger}\lf(-i\vect{\al}\cdot\vect{\nabla}+\beta
m_e\ri) \nu_e + \int d^3 \mathbf{x}
\,\nu_{\mu}^{\dagger}\lf(-i\vect{\al}\cdot\vect{\nabla}+\beta m_{\mu}\ri)
\nu_{\mu}
\\  \label{Hgauge}
&\equiv &\widetilde{H}_e(x_0) + \widetilde{H}_{\mu}(x_0).\eea 
It is worth stressing that the properties of the flavor fields are computed non-perturbatively
and the flavor states remain eigenstates of $\widetilde{H}(x_0)$ whatever is the value of
$g$, which is assumed constant in~\cite{Blasone:2010zn}.
From Eq.~\ref{Hgauge}, it also follows that $ \widetilde{H}(x_0)$ is
invariant for the $\nue \leftrightarrow \numu $ symmetry if (and only
if) $ m_e = m_{\mu}$, i.e. for $\theta \rightarrow \pi/4$.

\noindent
We are finally able to restate Conjecture (A) as a conjecture on the
coupling strength of the field $A_\mu$:

\noindent
{\bf Conjecture (B)} The mixing field strength $g$ is a function of
the number of flavor states $N$ that can potentially be observed through
the production of the corresponding charged leptons. In the limit $N
\rightarrow N_f$, $g \rightarrow 0$. If $N<N_f$, $g$ is settled to
restore flavor invariance for the Hamiltonian $\widetilde{H}(x_0)$.

Here, $N_f$ is a generic number of flavors but we remind that the derivation
of Eqs.~\ref{eq:euler-lagrange}-\ref{Hgauge} is done in two-flavor
approximation, so that quantitative predictions can be drawn only for
$N_f=2$. Note also that the requirement that the flavor states have to
be potentially observable through their projectors, i.e. their
capability of producing final state leptons in CC interactions, is
explicitly linked to $\widetilde{H}(x_0)$. This is in agreement with
the interpretation of $\widetilde{H}(x_0)$ as ``the energy that can be
extracted from flavor neutrinos through scattering'' given
in~\cite{Blasone:2010zn}.

\section{Facing experimental data}
\label{sec:data}

Although Conjecture (B) is stated in a quite rigorous manner, its
predictivity is limited by the underlying assumption of two-family
mixing (see Eq.\ref{PontecorvoMix}). A precise comparison with current
oscillation data necessarily requires an extension of the theory up to
the realistic $N=3$ case. Still, some information can already be drawn
for oscillation data, especially at the atmospheric scale. In this
regime, most of the experiments run at $E_\nu \gg m_\mu$, i.e. within
an energy range where the only unobservable flavor state is \nutau.
The three notably exceptions are the long-baseline reactor experiments
CHOOZ~\cite{chooz} and Palo Verde~\cite{palo_verde} ($N=1$), the
long-baseline accelerator experiment OPERA~\cite{OPERA} ($N=3$) and
the neutrino telescopes Ice Cube~\cite{icecube}, NEMO~\cite{nemo} and
ANTARES~\cite{antares} ($N=3$). All other experiments (K2K~\cite{k2k},
MINOS~\cite{minos}, SuperKamiokande I-III~\cite{superk}) operating at
the peak of oscillation probability for \dmtt\ work at $m_\mu \ll
E_\nu \ll 2m_\tau$, very far from the kinematic thresholds for muon
($\simeq m_\mu$) and tau ($\simeq 2m_\tau$) production.

Conjecture (B) therefore suggests that the mixing that is fully
operative in this region is the mixing toward the unobservable state
\nutau, while we can neglect transitions toward \nue\ that are
suppressed by $ E_\nu \gg m_e$. In the standard three-family
interpretation, it corresponds to $\thetatt \rightarrow \pi/4$ and
$\thetaot \rightarrow 0$, which is clearly in agreement with
experimental data. Hoverer, unlike the standard PMNS theory where
$\thetatt$ is universal except for negligible RGE
effects~\cite{running}, Conjecture (B) suggests that the measurement
of $\thetatt$ by MINOS and SuperKamiokande will differ from $\thetatt$
measured by OPERA and the neutrino telescopes, being $\thetatt^{N=2} >
\thetatt^{N=3}$.  OPERA~\cite{opera_tau} and the SuperKamiokande tau
appearance analysis~\cite{superk_tau} test in a direct manner the
appearance of tau neutrinos but due to the limited statistics they
cannot perform a precise measurement of $\thetatt$. Conjecture (B),
however, anticipates a fading of the mixing for $E \gg 2m_{\tau}$,
which in turn implies a dumping of the oscillations even in \numu
disappearance mode. Unfortunately, a test of Conjecture (B) in
disappearance mode is difficult with current facilities.  For neutrino
energies of a few tens of GeV the oscillation length is comparable or
larger than the earth radius and the flux of atmospheric neutrinos is
strongly suppressed. SuperKamiokande (SK) has selected high energy
samples in the SK-I,II and III data taking. However the sample of
``upward showering through-going muon events''~\cite{sk_thesis} has an
energy that is too large to exhibit oscillations even at baselines
comparable with the earth diameter. Some evidence for \numu
disappearance has been gained using the sample of ``upward
non-showering through-going muon events''~\cite{sk_thesis} but, again,
the disappearance is dominated by the low energy tail of the spectrum
($E<10$~GeV) and a precision measurement of $\thetatt$ is out of
reach. Similar considerations hold for the past MACRO~\cite{macro}
experiment at LNGS, for the recent high energy analysis of
SNO~\cite{SNO_highe} and for OPERA, which is missing a near detector
to perform disappearance searches. Finally, neutrino telescopes have a
typical muon threshold $\gg 100$~GeV, so that \numu disappearance is
unobservable both in the standard PMNS theory and under Conjecture
(B). A facility that might be capable of a conclusive test of
Conjecture (B) would be a neutrino telescope with an energy threshold
of ${\cal O}$(10) GeV. In fact, the Ice Cube DeepCore
array~\cite{deepcore_det} has been built to lower the threshold of Ice
Cube~\cite{icecube} down to 10~GeV and a very clear oscillation dip is
expected at an energy of
$\sim$25~GeV~\cite{deepcore_phys,FernandezMartinez:2010am,Mena:2008rh} in the
standard theory. Conjecture (B) anticipates that the $\nu_\mu$ deficit observed by
DeepCore will be much smaller than the one predicted by the PMNS theory.
  
Moving down toward $m_\mu$, the theory becomes non-predictive since it
cannot account for the interplay of the unobservable states \numu and
\nutau, especially for the disappearance of \nue. In particular, a
full three-flavor model is needed to explain why the leading angle
that determines $\nue$ disappearance of
solar~\cite{Homestake,Gallex-GNO,Sage,SK-solar} and very-long baseline
reactor neutrinos~\cite{kamland,borexino} is not exactly maximal
($\simeq$33$^\circ$ vs 45$^\circ$). Similarly, a naive two-family
approximation cannot be used to study the CHOOZ and Palo Verde
results, which have the same number of kinematic thresholds as KAMLAND
($N=1$) but run at the peak of \dmtt. It is, however, worth noticing
that the large number of experiments that are going to search for a
non-zero value of \thetaot~\cite{Mezzetto:2010zi,Battiston:2009ux}
will run at very different thresholds:
Double-Chooz~\cite{DoubleChooz}, Daya-Bay~\cite{DayaBay} and
RENO~\cite{RENO} at $N=1$ and $E_\nu \ll m_\mu$;
T2K~\cite{T2K} quite far from the muon production threshold
($E_\nu \simeq 6 m_\mu$ and $N=2$); MINOS and NOVA~\cite{NOVA} at
$E_\nu \gg m_\mu$ and $N=2$; OPERA in \nue appearance
mode~\cite{Komatsu:2002sz,migliozzi} at $E_\nu \gg 2m_\tau$ and $N=3$.
Again, inconsistent results between \thetaot measured by reactors,
T2K, MINOS/NOVA and OPERA would be a clear demonstration of the
non-universality of the PMNS and, possibly, of the correctness of
Conjecture (B). Table \ref{tab:data} summarizes these considerations,
showing the experiments that run near the peak of the oscillation
probability at the atmospheric and solar scale. Null results from past
experiments running far from the peak (CHORUS, NOMAD, CDHS, Bugey4
etc.~\cite{PDG}) do not add significant information about Conjecture
(B) and are not included. Note also that, in its present form,
Conjecture (B) does not anticipate any significant effect neither at
LSND~\cite{lsnd} nor at
MiniBoone~\cite{AguilarArevalo:2008rc,AguilarArevalo:2010wv}~\footnote{These
experiments, which are also included in Tab.~\ref{tab:data}, run at
different values of $N$: $N=1$ for LSND and $N=2$ for Miniboone.}.
  
Finally, it is worth mentioning that Conjecture (B) automatically preserves
the rate of neutral current interactions; it hence requires
that no NC deficit is observed for any value of $N$. At present, this
statement is in agreement with experimental data~\cite{PDG}.

\begin{sidewaystable}[ht]
\centering
\begin{tabular}{c|c|c|c|c|c|c|c}
\hline \hline 
Experiment & $E_\nu$ & e thresh. & $\mu$ thresh. & $\tau$ thresh. & peak & angle & expectation  \\ \hline
DeepCore~\cite{deepcore_phys} & $\sim25$ GeV & yes & yes & yes & $\sim \dmtt$  & $\thetatt$ & small \\
OPERA~\cite{opera_tau} & 17 GeV & yes & yes & yes & $\sim \dmtt$ (off-peak) & $\thetatt$ & small \\
MINOS~\cite{minos} &  3 GeV & yes & yes & no & $\dmtt$ &  $\thetatt$ & $\sim 45^\circ$ \\
SuperK (MG)~\cite{superk} & $\sim 3$~GeV & yes & yes & no & $\sim \dmtt$ &  $\thetatt$ & $\sim 45^\circ$ \\
K2K~\cite{k2k} &  1.3 GeV & yes & yes & no & $\dmtt$ &  $\thetatt$ & $\sim 45^\circ$ \\
T2K~\cite{T2K} & 600 MeV  & yes & yes & no & $\dmtt$ &  $\thetaot$ & small \\
Miniboone~\cite{AguilarArevalo:2008rc} & 600 MeV & yes & yes & no & $\sim 1$~eV$^2$ (off-peak)  & unknown & unknown \\ 
LSND~\cite{lsnd} & 30 MeV  & yes & no & no & $\sim 1$~eV$^2$ (off-peak)  & unknown & unknown \\ 
CHOOZ~\cite{chooz} & 3 MeV & yes & no & no &  $\dmtt$ &  $\thetaot$ & unknown \\
KAMLAND~\cite{kamland} & 3 MeV & yes & no & no &  $\dmot$ &   $\thetatt$ & unknown \\
Borexino~\cite{borexino} & 0.8 MeV & yes & no & no &  $\dmot$ &   $\thetaotw$ & unknown \\ 
SNO CC~\cite{SNO} & 8 MeV & yes & no & no &  $\dmot$ &   $\thetaotw$ & unknown \\ 
SuperK solar~\cite{SK-solar} & 8 MeV & yes & no & no &  $\dmot$ &   $\thetaotw$ & unknown \\ 
GNO-SAGE~\cite{Gallex-GNO,Sage} & 0.3 MeV & yes & no & no &  $\dmot$ &   $\thetaotw$ & unknown \\
\hline \hline 

\end{tabular}
\label{tab:data}
\caption{Summary of experimental data in the proximity of the
oscillation peaks or where appearance results have been obtained
(LSND, Miniboone, OPERA).  $E_\nu$ is the approximate neutrino energy
(see the corresponding references for details); the open kinematic
thresholds that are available to each experiment are shown in the
``e,$\mu$,$\tau$ thresh.''  columns. The $\Delta m^2$ probed by the
L/E of the experiment is indicated in the column labeled ``peak''; the leading mixing angle (in the PMNS interpretation)
is shown in the column ``angle'' together with the expectation from Conjecture (B)
(``expectation''). SuperK (MG) is the Multi-GeV analysis of
SuperKamiokande, while SNO CC represents the $\nue d \rightarrow p
p e$ analysis of SNO (phase I-II-III).  GNO-SAGE are the data
from the GALLEX-GNO and SAGE experiments.}

\end{sidewaystable}

\section{Conclusions}
\label{sec:conclusions}

In this paper we discussed the possibility that lepton mixing
originates from the low energy behaviour of interacting fermion
fields.  In this framework, mixing is a process dependent phenomenon,
whose size depends on the number of flavor states that can 
potentially be observed through the production of the corresponding
lepton. Running from below to above the kinematic thresholds, the
mixing parameters are expected to change and in the limit $E_\nu \ll
E^{thr}$ they settle to restore flavor invariance for the appropriate
Hamiltonian.  Employing and re-interpreting the results
of~\cite{Blasone:2010zn}, we were able to write explicitly such
Hamiltonian at least in two-family approximation and determine its
eigenstates also in the non-perturbative domain of the theory. This
conjecture, which is consistent with the seemingly bi-trimaximal
pattern~\cite{Harrison:2002er} of the PMNS, predicts non-universality
of the PMNS itself at scales much smaller than the electroweak scale;
it also anticipates a difference between the mixing angles that will
be measured by experiments running at different open thresholds
(different $N$). Notably, we expect that reactor experiments, T2K,
MINOS-NOVA and OPERA will measure different values of
\thetaot. Similar considerations hold for \thetatt as measured by
MINOS/SuperKamiokande and OPERA. Finally, the conjecture can be tested
in a direct manner through disappearance of \numu with energies $\gg$
$2m_{\tau}$. It therefore predicts that the neutrino deficit observed
by the Ice Cube DeepCore array will be significantly smaller than the
one of the standard PMNS theory. DeepCore will thus assess soon
whether the hypothesis that has been put forward here is a viable
option to explain the origin of leptonic mixing.

\section*{Acknowledgments}
Discussions with R.~Felici, C.~Giunti, F.~Halzen, E.~Lisi, P.~Migliozzi, D.~Nicol\`o and F.~Vissani
are gratefully acknowledged.  It is a great pleasure to thank
M.~Blasone, A.~Capolupo and G.~Vitiello for many useful insights on the physics of
flavor neutrino states.



\begin{thebibliography}{999}
\bibitem{why_nearly} The sets are not completely disconnected due to
  the momentum spread of initial particles.
\end{thebibliography}
\end{document}